\documentclass[runningheads]{llncs}

\makeatletter
\newcommand{\@chapapp}{\relax}%
\makeatother
\usepackage[T1]{fontenc}
\usepackage{graphicx}
\usepackage{amsfonts}
\usepackage{amsmath}
\usepackage{hyperref}
\usepackage{longtable}
\usepackage[title,header]{appendix}
\usepackage{caption}
\begin{document}
\title{Testing a cellular automata construction method to obtain 9-variable cryptographic Boolean functions\thanks{This work has been supported by a government grant managed by the Agence Nationale de la Recherche under the Investissement d'avenir program, reference ANR-17-EURE-004}}
\titlerunning{Testing a CA construction method to obtain 9-var crypto BF}
%
\author{Thomas Prévost\inst{1}\orcidID{0009-0000-2224-8574} \and
Bruno Martin\inst{1}\orcidID{0000-0002-0048-5197}}
\authorrunning{T. Prévost et al.}
%
\institute{Université Côte d'Azur, CNRS, I3S, France \\
\email{\{thomas.prevost,bruno.martin\}@univ-cotedazur.fr}}
\maketitle              
\begin{abstract}
We propose a method for constructing 9-variable cryptographic Boolean functions from the iterates of 5-variable cellular automata rules. We then analyze, for important cryptographic properties of 5-variable cellular automata rules, how they are preserved after extension to 9-variable Boolean functions. For each cryptographic property, we analyze the proportion of 5-variable cellular automata rules that preserve it for each of the 48 affine equivalence classes.

\keywords{Boolean functions \and Cellular automata \and Cryptography \and Affine equivalence classes.}
\end{abstract}

\section*{\uppercase{Introduction}}

Boolean functions play a dominant role in many cryptographic primitives. They are particularly used in hash functions~\cite{megha2016hash,daemen1993framework} or even symmetric block encryption~\cite{wu2016boolean}. These functions take a certain number of variables as input to return a unique Boolean binary value. Cellular automata rules can be considered as Boolean functions.

Some cellular automata rules have interesting cryptographic properties, notably for generating a pseudo random or chaotic output relative to the input passed to them. These rules can produce an output that is non-linear and statistically independent of the bits passed to them as input for example. They can be used for cryptographic applications, such as hashing or block encryption. Using these rules then avoids known attacks against cryptographic primitives, such as linear cryptanalysis~\cite{biryukov2011linear} for example. The first study of these chaotic functions was made by Wolfram in 1983, who discovered rule 30 with 3 variables~\cite{wolfram1983statistical}. Since then, numerous classifications of Boolean functions have been proposed~\cite{tsai1997boolean,braeken2004classification}. Many scientific papers study the use of Boolean functions in cryptography~\cite{formenti2014advances}. In particular, Boolean functions are used in cellular automata to construct hash functions~\cite{jamil2012new,hanin2017cahash,zheng1993haval}, or stream and block ciphers~\cite{nandi1994theory,kumaresan2017analytical}.

Finding good cryptographic Boolean functions from an exhaustive search becomes very difficult beyond 5 input variables. Indeed, the number of possible Boolean functions for $n$ input variables is $2^{2^n}$. If there are 4 294 967 296 possible 5-variable Boolean functions, there are on the other hand $1.34\cdot10^{154}$ 9-variable functions. It is therefore impossible, with current computing capabilities, to study them all exhaustively within a reasonable time. This is why we are looking for new methods to discover 9-variable Boolean functions with interesting cryptographic properties without studying them all exhaustively. We thus explore a method based on the evolution of a uniform cellular automaton to extend 5-variable cellular automata rules towards Boolean functions with 9 variables.

This work can be seen as an extension of the paper ``\emph{Advances on Random Sequence Generation by Uniform Cellular Automata}''~\cite{formenti2014advances}, which had already classified the conservation of some cryptographic properties after the extension of 5-variable cellular automata rules to 9-variable Boolean functions. In this paper, we exhaustively analyze the set of $2^{32}$ 5-variable cellular automata rules, study more cryptographic properties, and propose a more complete classification of the rules according to their affine equivalence classes.

First, we will introduce uniform cellular automata. Then we will talk about Boolean functions, their equivalence classes, and we will discuss their most important cryptographic properties. We will analyze, by affine equivalence class, the set of 5-variable cellular automata rules, and determine which ones have good cryptographic properties. Next we will explain our methodology for extending 5-variable cellular automata rules to 9-variable Boolean functions using a uniform cellular automaton. We will then discover to what extent the different cryptographic properties are preserved depending on the affine equivalence classes when extending the Boolean functions. We will show that this method is not very adequate to find 9-variable Boolean functions with good cryptographic properties since except for some properties like algebraic degree or balancedness of some affine equivalence classes, this method is not efficient to preserve cryptographic properties in extended Boolean functions.

\section{\uppercase{Uniform cellular automata}}

\begin{figure}
\centering
\includegraphics[width=0.8\textwidth]{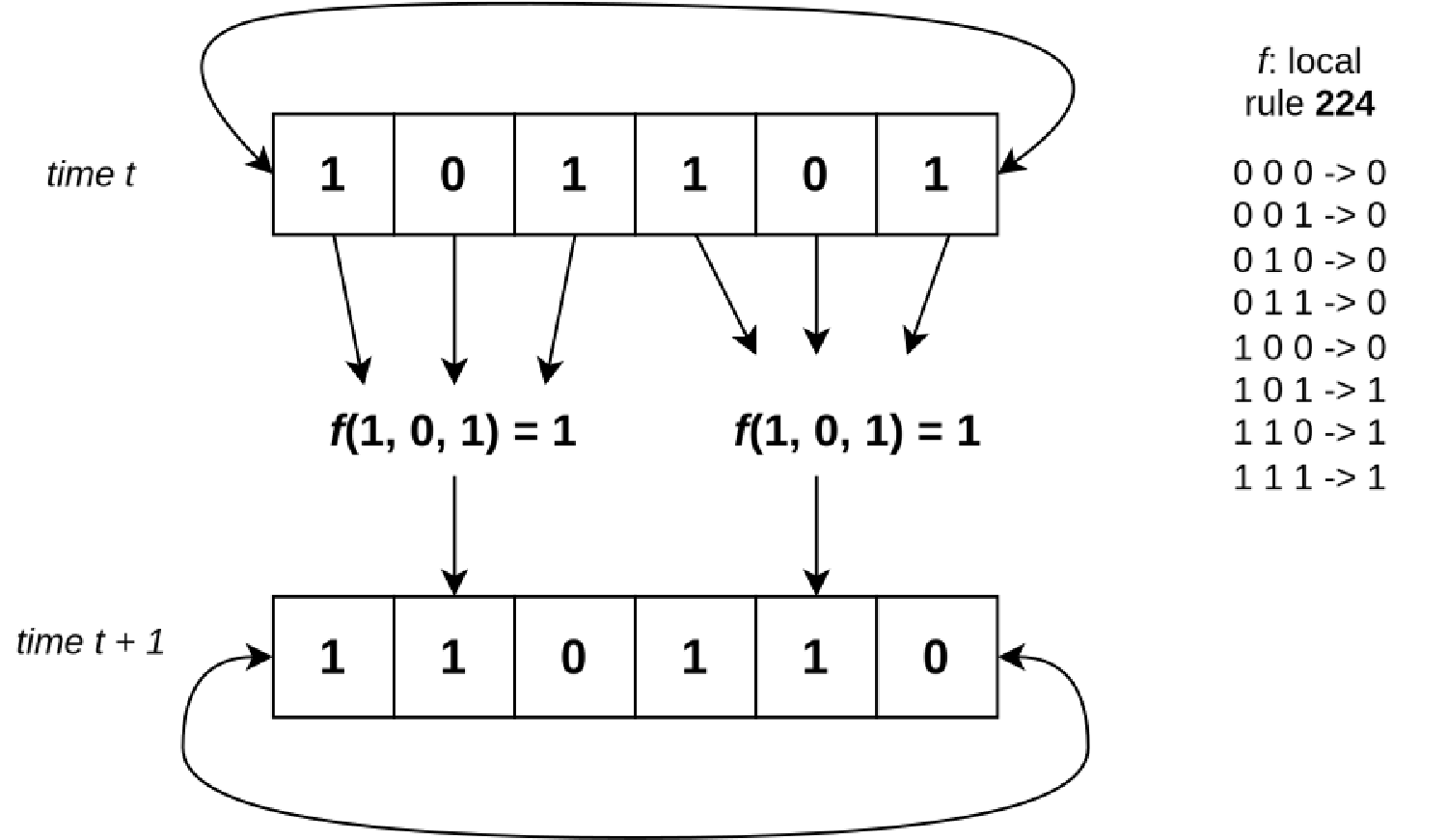}
\caption{A 1-dimensional uniform cellular automaton with the single 3-bit local rule 224. 224 is the binary representation of the rule's truth table.} \label{fig_simple_ca}
\end{figure}

A Cellular Automaton (CA) is a model of discrete parallel computation. It is composed of cells, each containing a unique Boolean value. For each time step, cells of the automaton are updated synchronously according to their value and the values of the neighboring cells, following a local rule. The local rule can be represented as a Boolean function, with multiple Boolean values as input (the cell and its neighbors) and a unique output value (the final cell value). An automaton is said to be \emph{uniform} if there is only one local rule for all the cells. Fig.~\ref{fig_simple_ca} shows an example of a uniform cellular automaton.

Here we study one-dimensional cellular automata. Formally, we can define them as triples $(Q, \delta, N)$ where
\begin{itemize}
    \item $Q$ is a finite set of states, here the set of states is $\{0, 1\}$, the Boolean values.
    \item $N \subseteq \mathbb{Z}$ is the finite neighborhood, $card(N)$ being the size of the CA.
    \item $\delta:Q^{card(N)} \longrightarrow Q$ is the local transition function, called \emph{rule}. $card(N)$ is the \emph{arity} of the rule. Here we use Boolean functions as local transition rules.
\end{itemize}

\section{\uppercase{Cryptographic properties of Boolean functions}}

\subsection{Representation of Boolean functions} \label{bool_f_repr}

Boolean functions are usually represented by their truth table, with the output of the function with all input bits set being the high order bit, and the output with all input bits unset being the low order bit.

\begin{example}
    The 3-variable rule 110 is written $0 1 1 0 1 1 1 0$ in
     binary. Starting with the rightmost significant bit, this means that
     $f(0, 0, 0) = 0$, $f(0, 0, 1) = 1$, $f(0, 1, 0) = 1$ etc.
\end{example}

We often use the hexadecimal representation of the truth table. The hexadecimal representation of rule 110 is $6e$. The hexadecimal representation will be extended to 5 and 9 variable Boolean functions later.

Boolean functions can also be expressed by their Algebraic Normal Form (ANF).

\begin{definition} \label{anf_def}
    Any $n$-variable Boolean function $f$ can be expressed by a unique binary polynomial,
     its Algebraic Normal Form (ANF):

     $f(x) = \bigoplus_{u\in \mathbb{F}^n_2}a_u(\prod^n_{i=1}x_i^{u_i}), a_u\in\mathbb{F}_2, u_i$ being the i-th projection of u, $x_i$ being the i-th bit of input $x \in \mathbb{F}_2$.
\end{definition}

\begin{example}
    The Algebraic Normal Form of 3-variable rule 110 is $f(x) = f(x_0, x_1, x_2) = x_0x_1x_2 + x_0x_1 + x_0 + x_1$.
\end{example}

\subsection{Affine equivalence classes}

It is possible to classify the set of $2^{2^n}$ $n$-variable Boolean functions into different equivalence classes. This often makes it easier to study these functions, by potentially reducing the set of functions to be studied.

\begin{definition}
    Two $n$-variable Boolean functions $f$ and $g$ are said to be \emph{affine equivalent} if there exists a singular $n \times n$  Boolean square matrix $A$, two vectors $b, c \in \mathbb{F}_2^n$ and a boolean value $d \in \mathbb{F}_2$ such that $\forall x \in \mathbb{F}_2^n$
    \begin{equation}
        f(x) = g(Ax + b) + c \cdot x + d
    \end{equation}
\end{definition}

An affine equivalence class groups together all functions that are affinely equivalent to each other.

Different algorithms have been proposed to determine whether two $n$-variable Boolean functions belong to the same affine equivalence class~\cite{meng12005analysis,wang2019method}. \cite{zeng2019computing} gives us the number of affine equivalence classes for $n$-variable Boolean functions, up to $n = 9$. We thus know that there are 48 affine equivalence classes of 5-variable Boolean functions. Table~\ref{table_eq_classes}, in \ref{appendice_eq_classes}, gives the list of affine equivalence classes for 5-variable Boolean functions, with a representative function for each class, its ANF form as well as the total number of 5-variable Boolean functions in this class.

In the remainder of this paper, we will use the representative function presented in table~\ref{table_eq_classes} to denote the corresponding affine equivalence class. For example, ``the equivalence class $aa55aa55$'' means ``the set of rules which are affine equivalent to $aa55aa55$''.

\subsection{Balancedness}

The balancedness property is important for the choice of cryptographic Boolean functions. This characteristic is essential for achieving uniformity, which prevents bias and makes it more difficult for attackers to distinguish between the possible outputs~\cite{carlet2010boolean}.

\begin{definition}
    A Boolean function $f$ is balanced if and only if it returns true for 50\% of possible inputs.
\end{definition}

\subsection{Correlation immunity}

Correlation immunity is important for cryptographic Boolean functions because it ensures that the function's output is statistically independent of any subset of its input variables. This property protects against Siegenthaler's correlation attacks, where an attacker attempts to exploit linear relationships between the inputs and outputs to uncover secret information~\cite{Carlet2011}.

\begin{definition} \label{correlation_immune_def}
    An $n$-variable Boolean function $f$ is k-order correlation immune, $1 \le k \le n$, if and only
     if $f(x)$ is statistically independent from any subset of size $k$ of $x$, $x = x_1, ..., x_n$
     being any binary random input.
\end{definition}

The Walsh-Hadamard transform helps us know if a function is correlation-immune. The Walsh-Hadamard
 transform of a Boolean function $f$ is given by:

\begin{equation}
    S_f(\omega) = \sum_{x=0}^{2^n-1}(-1)^{f(x) \oplus x \cdot \omega}
\end{equation}
where $x \cdot \omega = \sum_{i=0}^{n-1}x_i \cdot w_i$ is the dot product of the two binary vectors.

\begin{theorem}
    A $n$-variable Boolean function $f$ is k-order correlation immune, $1 \le k \le n$ if and only if for every $\omega \in \mathbb{F}_2^n$ such that $1 \le w_h(\omega) \le k$, $S_f(\omega) = 0$.
\end{theorem}

With $w_h(\omega)$ the \emph{Hamming weight} of the function $\omega$, i.e. the number of entries $x \in \mathbb{F}_2^n$ such that $\omega(x) = 1$. This theorem has been proved by Xiao and Massey in~\cite{xiao1988spectral}.

A correlation immune to order $k$ and balanced function is said to be \emph{resilient to order $k$}.

\subsection{Strict avalanche criterion}

\begin{definition} \label{sac_def}
    A $n$-variable Boolean function $f$ satisfies the Strict Avalanche Criterion (SAC) if
     $\forall i \in [\![1, n]\!]$, flipping the i-th bit of the input x results in the output
     $f(x)$ being changed for exactly 50\% of the inputs $x$.
\end{definition}

Satisfying the strict avalanche criterion makes the Boolean function ``chaotic'', and thus makes
 difficult to infer its input from its output. It is particularly interesting for cryptographic
 applications~\cite{upadhyay2022investigating}.

\subsection{Propagation criterion}

\begin{definition}
    An $n$-variable Boolean function $f$ is said to satisfy the propagation criterion at order $k$ if
     its output changes with a propability of 50\% when we invert $i$ bits of its input,
     $\forall i \in [\![1, k]\!]$.
\end{definition}

The propagation criterion property is crucial for achieving high levels of diffusion, which means
 that small changes in the input spread out and affect many output bits. High diffusion makes it
 harder for attackers to predict or control the output by manipulating the
 input~\cite{gouget2004propagation}.

Satisfying the first-order propagation criterion is equivalent to satisfying the strict avalanche
 criterion.

\subsection{Algebraic degree}

Any $n$-variable Boolean function can be expressed in its ANF form, as explained in definition~\ref{anf_def}. The algebraic degree of a function $f$ is the number of variable in the largest monomial $x_1^{u_1} ...\ x_n^{u_n}$ of its ANF.

A higher algebraic degree increases the function's resistance to algebraic attacks, where attackers try to solve or approximate the system of equations representing the cryptographic algorithm~\cite{millan1995low}. Functions with a high algebraic degree are more complex and harder to approximate.

A Boolean function is said to be \emph{nonlinear} if its algebraic degree at least 2. These functions are interesting for cryptography, for example they allow protection against linear cryptanalysis~\cite{matsui1993linear} if they are used to construct S-boxes.

\section{\uppercase{Analysis of the cryptographic properties of 5-variable Boolean functions}} \label{sect_5_var_analysis}

Before trying to extend the 5-variable cellular automata rules to 9-variable Boolean functions, we were interested in the 5-variable CA rules that validate good cryptographic properties. We therefore analyzed the set of 5-variable CA rules to find out which ones validate the following cryptographic properties:
\begin{itemize}
    \item Algebraic degree
    \item Linearity
    \item Strict avalanche criterion
    \item Balanced
    \item First-order correlation immunity
    \item Propagation criterion from degree 2 to 5.
\end{itemize}

We grouped the rules by affine equivalence class, the results can be found in \ref{5_variable_properties}.

The least interesting affine equivalence class for cryptographic use is class $aa55aa55$, which is the equivalence class of linear functions (algebraic degree $\leq$ 1).

Class $288d1b41$ is interesting because it encompasses many functions with interesting cryptographic properties. It is the only class to encompass functions that meet the propagation criterion at order 3 and 4.

Class $88ddbb11$ is also very interesting: although none of its functions meet the propagation criterion from order 2, 26\% of its functions meet the Strict Avalanche Criterion, 49\% meet the first order Correlation Immunity, and 88\% are balanced.

Classes $288d1b41$ and $88ddbb11$ are only of degree 2, but the algebraic degree is not preserved when extending into a Boolean function with 9 variables. Rules belonging to these classes could therefore be interesting to be extended with our method.

\section{\uppercase{Our method to extend Boolean functions from 5 to 9 variables}}

In order to facilitate the search for ``good'' 9-variable Boolean functions, we seek to obtain them by extension of 5-variable Boolean functions. We must also ensure that the cryptographic properties of the latter are preserved after extension.

\subsection{Iterates of a cellular automaton}

\begin{figure}
\includegraphics[width=\textwidth]{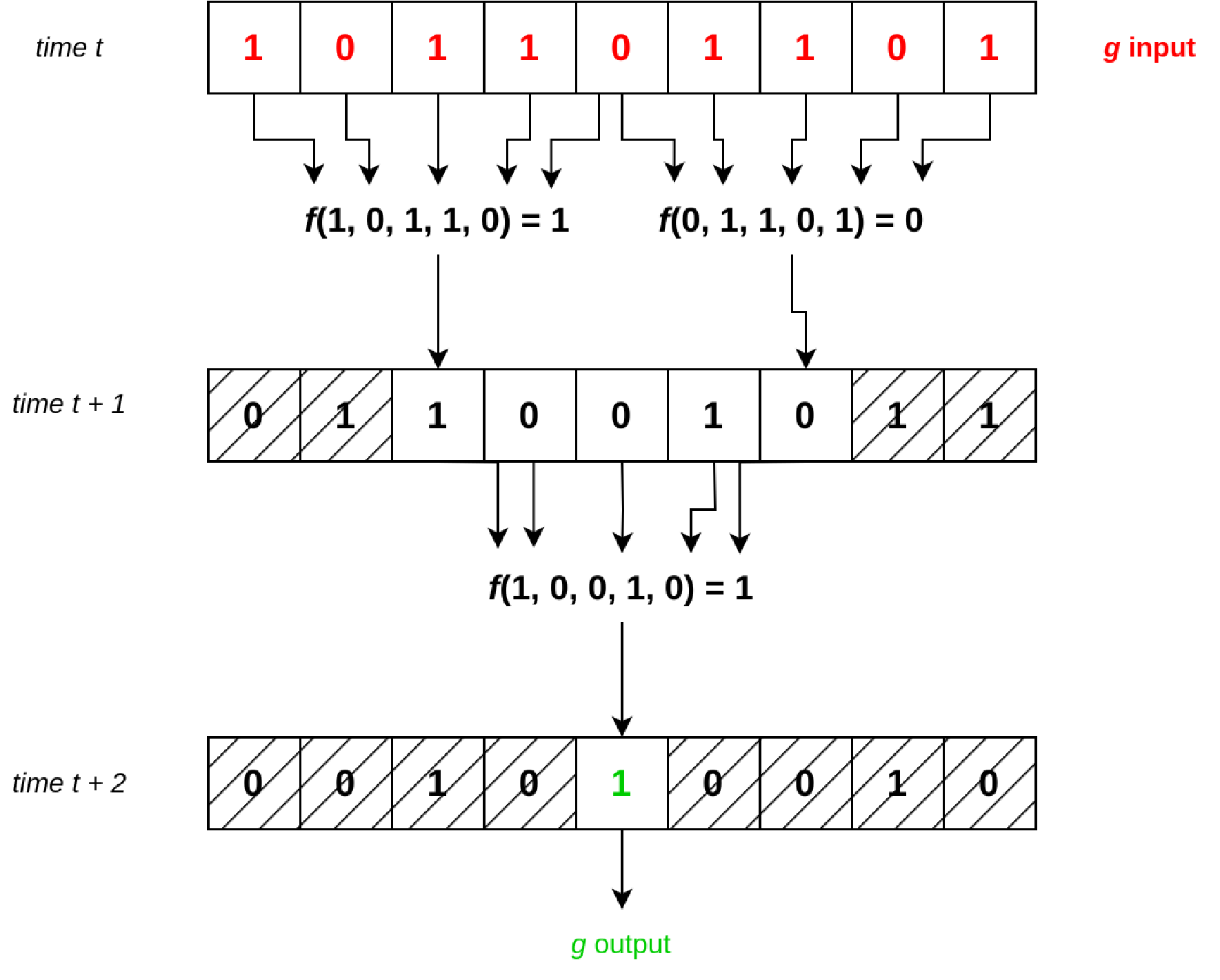}
\caption{Extension of the 5-variable cellular automata rule $f$ to the 9-variable Boolean function $g$. The 9 input bits of $g$ are placed in the cells of the uniform cellular automaton. We apply the local rule $f$ twice, then the value of the central cell of the cellular automaton is the output of the function $g$. Hatched cells are no longer necessary at the corresponding time step to calculate the output of $g$.} \label{fig_ring_5_to_9}
\end{figure}

In order to extend a 5-variable cellular automata rule $f: \{0, 1\}^5 \longrightarrow \{0, 1\}$ to a 9-variable Boolean function $g: \{0, 1\}^9 \longrightarrow \{0, 1\}$, we use a cellular automaton with 9 cells.

As shown in fig.~\ref{fig_ring_5_to_9}, we pass the input value of $g$ into the 9 cells of the cellular automaton. Then, we use the 5-variable cellular automata rule $f$ as the only local transition
 rule on the automaton. The cellular automaton is therefore uniform, with the local transition $f$.

We apply the local transition twice in a row, so that the values of each cell can be propagated throughout the automaton. Finally, we choose the value at the center of the automaton as the output of the function $g$.

We apply this process for all $2^9 = 512$ possible entries of $g$ (000000000, 000000001, etc.), we thus obtain as output the truth table of $g$.

\subsection{Testing the different properties of 9-variable Boolean functions}

The demonstration method we choose to validate the different properties of obtained 9-variable Boolean functions is quite simple: we test the entire domain of definition of 5-variable cellular automata rules. Indeed the set of $n$-variable cellular automata rules is finite, there are $2^{2^n}$ different functions. There are therefore $2^{2^5} = 4 \ 294 \ 967 \ 296$ 5-variable cellular automata rules, so it is possible to test them exhaustively with a modern CPU.

We tested the following properties on Boolean functions with 9 variables:
\begin{itemize}
    \item Algebraic degree
    \item Nonlinearity
    \item Strict avalanche criterion
    \item Balanced
    \item First-order correlation immunity
    \item Propagation criterion from degree 2 to 5.
\end{itemize}
 
We then compared these properties to those of associated 5-variable cellular automata rules.

We count, for each affine equivalence class, how many 9-variable Boolean functions preserve the property of the 5-variable rule. The property is said to be ``preserved'' if it holds for both the 5-variable CA rule and the extended 9-variable Boolean function. For example, if $f$ is a 5-variable rule, and $g$ is the extended 9-variable function associated, the property of nonlinearity is said to be preserved if $f$ and $g$ are both nonlinear.

The algebraic degree is considered to be ``preserved'' if its value does not change or increases after the extension, meaning $deg(g) \geq deg(f)$.

The code we used for our exhaustive testing can be viewed at the following address: \url{https://github.com/thomasarmel/boolean_function_extender}. We optimized the algorithms for searching cryptographic properties, and parallelized the exhaustive search on a machine with two Intel Xeon E5-2696 v2 2.50GHz 12-core 24-thread CPUs. This allowed us to analyze the entire set of 5-variable CA rules in a reasonable time.

\section{\uppercase{Results}}

We have exhaustively analyzed the conservation of cryptographic properties of 5-variable CA rules after extension to 9-variable Boolean functions. The results are given in \ref{preserved_properties}.

Cryptographic properties are globally poorly preserved. However, it should be noted that extended Boolean functions generally retain an algebraic degree greater than or equal to the original 5-variable rule (which is not surprising because there are more variables). Nonlinearity is therefore preserved for most rules, with the exception of the 64 rules of the $aa55aa55$ linear class, for which all 9-variable Boolean functions remain linear after extension.

Properties such as the Strict Avalanche Criterion or the Propagation Criterion are never preserved. These cryptographic properties were already quite rare in 5-variable CA rules. Interestingly, the 62 rules of degree strictly equal to 1 remain balanced after extension to 9-variable Boolean functions.

Apart from the $aa55aa55$ class of linear rules, it is the functions of the affine equivalence class $88ddbb11$ ($x_0x_1 + x_0x_3 + x_0 + x_1x_4 + x_1 + x_3x_4 + 1$) that best preserves the properties of balancedness and first order correlation immunity after extension to 9-variable Boolean functions.

\section*{\uppercase{Conclusion}}
In this paper, we explored a method to generate 9-variable Boolean functions from 5-variable cellular automata rules, extended using iterations over a uniform cellular automaton. We found that cryptographic properties are generally poorly preserved by this extension method. Some properties are never preserved, such as the Strict Avalanche Criterion or the Correlation Immunity. On the other hand, nonlinearity and high algebraic degree are very well preserved by this method.

We also performed an analysis of the preservation of cryptographic properties according to the affine equivalence classes of 5-variable CA rules. We found that while first order Correlation Immunity and balancedness are generally poorly preserved after the extension to 9-variable Boolean functions, rules in certain equivalence classes such as $88ddbb11$ frequently extend to balanced 9-variable Boolean functions, or functions fulfilling the conditions of first order correlation immunity.

This extension method is therefore not very efficient in preserving the cryptographic properties of Boolean functions, apart from the algebraic degree, and the balancedness and first order correlation immunity for some affine equivalence classes. It would then be interesting to look for other methods using cellular automata to extend Boolean functions capable of better preserving the cryptographic properties, such as genetic algorithms for instance.

\subsubsection*{Acknowledgements} We would like to thank Jean-Charles Regin (\url{http://www.constraint-programming.com/people/regin/}) for kindly lending us his computing machines.

%
%
%
%
\bibliographystyle{splncs04}
\bibliography{extending_bool_5_to_9}

\begin{appendices}
\renewcommand{\thesection}{\appendixname~\Alph{section}}
\section{List of affine equivalence classes of 5-variable Boolean functions} \label{appendice_eq_classes}

\begin{longtable}{ |p{0.15\textwidth}|p{0.8\textwidth}|p{0.15\textwidth}| } 
    \captionsetup{width=1.1\linewidth}
    \hline
    \textbf{Class representative (hex)} & \textbf{Algebraic Normal Form} & \textbf{Functions count} \\
    \hline
    aa55aa55 & $x_0 + x_3 + 1$ & 64 \\
    \hline
    aa55ab55 & $x_0x_1x_2x_3x_4 + x_0x_1x_2x_3 + x_0x_1x_3x_4 + x_0x_1x_3 + x_0x_2x_3x_4 + x_0x_2x_3 + x_0x_3x_4 + x_0x_3 + x_0 + x_1x_2x_3x_4 + x_1x_2x_3 + x_1x_3x_4 + x_1x_3 + x_2x_3x_4 + x_2x_3 + x_3x_4 + 1$ & 2048 \\
    \hline
    aa55bb55 & $x_0x_1x_3x_4 + x_0x_1x_3 + x_0x_3x_4 + x_0x_3 + x_0 + x_1x_3x_4 + x_1x_3 + x_3x_4 + 1$ & 31744 \\
    \hline
    aa5dbb55 & $x_0x_1x_2x_3x_4 + x_0x_1x_2x_4 + x_0x_1x_3 + x_0x_1x_4 + x_0x_3x_4 + x_0x_3 + x_0 + x_1x_3x_4 + x_1x_3 + x_3x_4 + 1$ & 317440 \\
    \hline
    aaddbb55 & $x_0x_1x_3 + x_0x_1x_4 + x_0x_3x_4 + x_0x_3 + x_0 + x_1x_3x_4 + x_1x_3 + x_3x_4 + 1$ & 79360 \\
    \hline
    aa5dbb51 & $x_0x_1x_2x_3 + x_0x_1x_2 + x_0x_1x_3x_4 + x_0x_1 + x_0x_3x_4 + x_0x_3 + x_0 + x_1x_2x_3x_4 + x_1x_2x_3 + x_1x_2x_4 + x_1x_2 + x_1x_4 + x_1 + x_3x_4 + 1$ & 2222080 \\
    \hline
    2a5dbb51 & $x_0x_1x_2x_3x_4 + x_0x_1x_2x_3 + x_0x_1x_2 + x_0x_1x_3x_4 + x_0x_1 + x_0x_3x_4 + x_0x_3 + x_0 + x_1x_2x_3x_4 + x_1x_2x_3 + x_1x_2x_4 + x_1x_2 + x_1x_4 + x_1 + x_3x_4 + 1$ & 10665984 \\
    \hline
    aaddbb51 & $x_0x_1x_2x_3x_4 + x_0x_1x_2x_3 + x_0x_1x_2x_4 + x_0x_1x_2 + x_0x_1x_3x_4 + x_0x_1 + x_0x_3x_4 + x_0x_3 + x_0 + x_1x_2x_3x_4 + x_1x_2x_3 + x_1x_2x_4 + x_1x_2 + x_1x_4 + x_1 + x_3x_4 + 1$ & 2222080 \\
    \hline
    2a5dbf51 & $x_0x_1x_2 + x_0x_1x_3 + x_0x_1 + x_0x_3x_4 + x_0x_3 + x_0 + x_1x_2x_4 + x_1x_2 + x_1x_3x_4 + x_1x_3 + x_1x_4 + x_1 + x_3x_4 + 1$ & 1777664 \\
    \hline
    6a5dbb51 & $x_0x_1x_2x_3 + x_0x_1x_2 + x_0x_1x_3x_4 + x_0x_1 + x_0x_3x_4 + x_0x_3 + x_0 + x_1x_2x_3 + x_1x_2x_4 + x_1x_2 + x_1x_4 + x_1 + x_3x_4 + 1$ & 28442624 \\
    \hline
    2addbb51 & $x_0x_1x_2x_3 + x_0x_1x_2x_4 + x_0x_1x_2 + x_0x_1x_3x_4 + x_0x_1 + x_0x_3x_4 + x_0x_3 + x_0 + x_1x_2x_3x_4 + x_1x_2x_3 + x_1x_2x_4 + x_1x_2 + x_1x_4 + x_1 + x_3x_4 + 1$ & 26664960 \\
    \hline
    a8ddbb51 & $x_0x_1x_2x_3 + x_0x_1x_2x_4 + x_0x_1x_2 + x_0x_1 + x_0x_2x_3x_4 + x_0x_3 + x_0 + x_1x_2x_3x_4 + x_1x_2x_3 + x_1x_2x_4 + x_1x_2 + x_1x_4 + x_1 + x_3x_4 + 1$ & 1111040 \\
    \hline
    aeddda51 & $x_0x_1x_2x_3x_4 + x_0x_1x_2x_4 + x_0x_1x_2 + x_0x_1x_3x_4 + x_0x_1x_3 + x_0x_1 + x_0 + x_1x_2x_3 + x_1x_2x_4 + x_1x_2 + x_1x_3 + x_1x_4 + x_1 + x_2x_3x_4 + x_2x_3 + x_3 + 1$ & 28442624 \\
    \hline
    0a5dbf51 & $x_0x_1x_2x_3x_4 + x_0x_1x_2 + x_0x_1x_3 + x_0x_1 + x_0x_2x_3x_4 + x_0x_3x_4 + x_0x_3 + x_0 + x_1x_2x_4 + x_1x_2 + x_1x_3x_4 + x_1x_3 + x_1x_4 + x_1 + x_3x_4 + 1$ & 17776640 \\
    \hline
    8addda51 & $x_0x_1x_2x_3x_4 + x_0x_1x_2x_4 + x_0x_1x_2 + x_0x_1x_3 + x_0x_1 + x_0x_2x_3x_4 + x_0 + x_1x_2x_3x_4 + x_1x_2x_3 + x_1x_2x_4 + x_1x_2 + x_1x_3x_4 + x_1x_3 + x_1x_4 + x_1 + x_2x_3x_4 + x_2x_3 + x_3 + 1$ & 142213120 \\
    \hline
    a8dd9b51 & $x_0x_1x_2x_3x_4 + x_0x_1x_2x_4 + x_0x_1x_2 + x_0x_1 + x_0x_2x_3 + x_0x_3 + x_0 + x_1x_2x_3x_4 + x_1x_2x_3 + x_1x_2x_4 + x_1x_2 + x_1x_4 + x_1 + x_3x_4 + 1$ & 26664960 \\
    \hline
    88ddbb51 & $x_0x_1x_2x_3x_4 + x_0x_1x_2x_3 + x_0x_1x_2x_4 + x_0x_1x_2 + x_0x_1 + x_0x_3 + x_0 + x_1x_2x_3x_4 + x_1x_2x_3 + x_1x_2x_4 + x_1x_2 + x_1x_4 + x_1 + x_3x_4 + 1$ & 317440 \\
    \hline
    88ddbb11 & $x_0x_1 + x_0x_3 + x_0 + x_1x_4 + x_1 + x_3x_4 + 1$ & 9920 \\
    \hline
    8c5dda51 & $x_0x_1x_2 + x_0x_1x_3 + x_0x_1 + x_0x_3x_4 + x_0 + x_1x_2x_3 + x_1x_2x_4 + x_1x_2 + x_1x_3 + x_1x_4 + x_1 + x_2x_3x_4 + x_2x_3 + x_3 + 1$ & 17776640 \\
    \hline
    a89d9b51 & $x_0x_1x_2 + x_0x_1 + x_0x_2x_3 + x_0x_3 + x_0 + x_1x_2x_3 + x_1x_2 + x_1x_4 + x_1 + x_3x_4 + 1$ & 1666560 \\
    \hline
    8eddda51 & $x_0x_1x_2x_4 + x_0x_1x_2 + x_0x_1x_3x_4 + x_0x_1x_3 + x_0x_1 + x_0x_2x_3x_4 + x_0 + x_1x_2x_3 + x_1x_2x_4 + x_1x_2 + x_1x_3 + x_1x_4 + x_1 + x_2x_3x_4 + x_2x_3 + x_3 + 1$ & 106659840 \\
    \hline
    aefdda51 & $x_0x_1x_2 + x_0x_1x_3x_4 + x_0x_1x_3 + x_0x_1 + x_0x_2x_3x_4 + x_0x_2x_4 + x_0 + x_1x_2x_3 + x_1x_2x_4 + x_1x_2 + x_1x_3 + x_1x_4 + x_1 + x_2x_3x_4 + x_2x_3 + x_3 + 1$ & 284426240 \\
    \hline
    025dbf51 & $x_0x_1x_2 + x_0x_1x_3x_4 + x_0x_1x_3 + x_0x_1 + x_0x_2x_3x_4 + x_0x_3x_4 + x_0x_3 + x_0 + x_1x_2x_4 + x_1x_2 + x_1x_3x_4 + x_1x_3 + x_1x_4 + x_1 + x_3x_4 + 1$ & 19998720 \\
    \hline
    88ddda51 & $x_0x_1x_2x_4 + x_0x_1x_2 + x_0x_1x_3x_4 + x_0x_1x_3 + x_0x_1 + x_0x_3x_4 + x_0 + x_1x_2x_3x_4 + x_1x_2x_3 + x_1x_2x_4 + x_1x_2 + x_1x_3x_4 + x_1x_3 + x_1x_4 + x_1 + x_2x_3x_4 + x_2x_3 + x_3 + 1$ & 213319680 \\
    \hline
    88dd9b51 & $x_0x_1x_2x_4 + x_0x_1x_2 + x_0x_1 + x_0x_2x_3x_4 + x_0x_2x_3 + x_0x_3 + x_0 + x_1x_2x_3x_4 + x_1x_2x_3 + x_1x_2x_4 + x_1x_2 + x_1x_4 + x_1 + x_3x_4 + 1$ & 3809280 \\
    \hline
    ceddda51 & $x_0x_1x_2x_3x_4 + x_0x_1x_2x_4 + x_0x_1x_2 + x_0x_1x_3x_4 + x_0x_1x_3 + x_0x_1 + x_0x_2x_3x_4 + x_0 + x_1x_2x_3x_4 + x_1x_2x_3 + x_1x_2x_4 + x_1x_2 + x_1x_3 + x_1x_4 + x_1 + x_2x_3x_4 + x_2x_3 + x_3 + 1$ & 426639360 \\
    \hline
    0eddda51 & $x_0x_1x_2x_3x_4 + x_0x_1x_2x_4 + x_0x_1x_2 + x_0x_1x_3x_4 + x_0x_1x_3 + x_0x_1 + x_0x_2x_3x_4 + x_0 + x_1x_2x_3 + x_1x_2x_4 + x_1x_2 + x_1x_3 + x_1x_4 + x_1 + x_2x_3x_4 + x_2x_3 + x_3 + 1$ & 142213120 \\
    \hline
    425dbf51 & $x_0x_1x_2x_3x_4 + x_0x_1x_2 + x_0x_1x_3x_4 + x_0x_1x_3 + x_0x_1 + x_0x_2x_3x_4 + x_0x_3x_4 + x_0x_3 + x_0 + x_1x_2x_3x_4 + x_1x_2x_4 + x_1x_2 + x_1x_3x_4 + x_1x_3 + x_1x_4 + x_1 + x_3x_4 + 1$ & 319979520 \\
    \hline
    8cddda51 & $x_0x_1x_2x_3x_4 + x_0x_1x_2x_4 + x_0x_1x_2 + x_0x_1x_3 + x_0x_1 + x_0x_3x_4 + x_0 + x_1x_2x_3 + x_1x_2x_4 + x_1x_2 + x_1x_3 + x_1x_4 + x_1 + x_2x_3x_4 + x_2x_3 + x_3 + 1$ & 426639360 \\
    \hline
    88dddb51 & $x_0x_1x_2x_3x_4 + x_0x_1x_2x_3 + x_0x_1x_2x_4 + x_0x_1x_2 + x_0x_1 + x_0x_2x_3x_4 + x_0x_2x_3 + x_0x_3 + x_0 + x_1x_2x_4 + x_1x_2 + x_1x_4 + x_1 + x_3x_4 + 1$ & 20316160 \\
    \hline
    289d9b51 & $x_0x_1x_2x_3x_4 + x_0x_1x_2 + x_0x_1 + x_0x_2x_3 + x_0x_3 + x_0 + x_1x_2x_3 + x_1x_2 + x_1x_4 + x_1 + x_3x_4 + 1$ & 26664960 \\
    \hline
    86fdda51 & $x_0x_1x_2 + x_0x_1x_3 + x_0x_1 + x_0x_2x_4 + x_0 + x_1x_2x_3 + x_1x_2x_4 + x_1x_2 + x_1x_3 + x_1x_4 + x_1 + x_2x_3x_4 + x_2x_3 + x_3 + 1$ & 26664960 \\
    \hline
    88dddb71 & $x_0x_1 + x_0x_2x_4 + x_0x_2 + x_0x_3 + x_0 + x_1x_2x_4 + x_1x_2 + x_1x_4 + x_1 + x_3x_4 + 1$ & 5079040 \\
    \hline
    cefdda51 & $x_0x_1x_2 + x_0x_1x_3x_4 + x_0x_1x_3 + x_0x_1 + x_0x_2x_4 + x_0 + x_1x_2x_3x_4 + x_1x_2x_3 + x_1x_2x_4 + x_1x_2 + x_1x_3 + x_1x_4 + x_1 + x_2x_3x_4 + x_2x_3 + x_3 + 1$ & 170655744 \\
    \hline
    0efdda51 & $x_0x_1x_2 + x_0x_1x_3x_4 + x_0x_1x_3 + x_0x_1 + x_0x_2x_4 + x_0 + x_1x_2x_3 + x_1x_2x_4 + x_1x_2 + x_1x_3 + x_1x_4 + x_1 + x_2x_3x_4 + x_2x_3 + x_3 + 1$ & 639959040 \\
    \hline
    288d9b51 & $x_0x_1x_2x_4 + x_0x_1x_2 + x_0x_1 + x_0x_2x_3x_4 + x_0x_2x_3 + x_0x_2x_4 + x_0x_3 + x_0 + x_1x_2x_3x_4 + x_1x_2x_3 + x_1x_2x_4 + x_1x_2 + x_1x_4 + x_1 + x_2x_3x_4 + x_2x_4 + x_3x_4 + 1$ & 26664960 \\
    \hline
    8cfdda51 & $x_0x_1x_2 + x_0x_1x_3 + x_0x_1 + x_0x_2x_3x_4 + x_0x_2x_4 + x_0x_3x_4 + x_0 + x_1x_2x_3 + x_1x_2x_4 + x_1x_2 + x_1x_3 + x_1x_4 + x_1 + x_2x_3x_4 + x_2x_3 + x_3 + 1$ & 426639360 \\
    \hline
    8cdddb51 & $x_0x_1x_2x_3 + x_0x_1x_2x_4 + x_0x_1x_2 + x_0x_1x_3x_4 + x_0x_1 + x_0x_2x_3x_4 + x_0x_2x_3 + x_0x_3 + x_0 + x_1x_2x_3x_4 + x_1x_2x_4 + x_1x_2 + x_1x_3x_4 + x_1x_4 + x_1 + x_3x_4 + 1$ & 35553280 \\
    \hline
    8ccdda51 & $x_0x_1x_2 + x_0x_1x_3 + x_0x_1 + x_0x_2x_3x_4 + x_0x_2x_4 + x_0x_3x_4 + x_0 + x_1x_2x_3x_4 + x_1x_2x_3 + x_1x_2 + x_1x_3 + x_1x_4 + x_1 + x_2x_3 + x_2x_4 + x_3 + 1$ & 79994880 \\
    \hline
    289d9b41 & $x_0x_1x_2x_3 + x_0x_1x_2x_4 + x_0x_1 + x_0x_2x_3x_4 + x_0x_2x_4 + x_0x_2 + x_0x_3 + x_0 + x_1x_2x_3x_4 + x_1x_2x_4 + x_1x_4 + x_1 + x_2x_3x_4 + x_2x_3 + x_2x_4 + x_2 + x_3x_4 + 1$ & 888832 \\
    \hline
    488ddb51 & $x_0x_1x_2x_3x_4 + x_0x_1x_2x_3 + x_0x_1x_2x_4 + x_0x_1x_2 + x_0x_1 + x_0x_2x_3 + x_0x_2x_4 + x_0x_3 + x_0 + x_1x_2x_3x_4 + x_1x_2x_4 + x_1x_2 + x_1x_4 + x_1 + x_2x_3x_4 + x_2x_4 + x_3x_4 + 1$ & 106659840 \\
    \hline
    ccfdda51 & $x_0x_1x_2x_3x_4 + x_0x_1x_2 + x_0x_1x_3 + x_0x_1 + x_0x_2x_3x_4 + x_0x_2x_4 + x_0x_3x_4 + x_0 + x_1x_2x_3x_4 + x_1x_2x_3 + x_1x_2x_4 + x_1x_2 + x_1x_3 + x_1x_4 + x_1 + x_2x_3x_4 + x_2x_3 + x_3 + 1$ & 341311488 \\
    \hline
    688d9b51 & $x_0x_1x_2x_3x_4 + x_0x_1x_2x_4 + x_0x_1x_2 + x_0x_1 + x_0x_2x_3x_4 + x_0x_2x_3 + x_0x_2x_4 + x_0x_3 + x_0 + x_1x_2x_3 + x_1x_2x_4 + x_1x_2 + x_1x_4 + x_1 + x_2x_3x_4 + x_2x_4 + x_3x_4 + 1$ & 106659840 \\
    \hline
    288d9b41 & $x_0x_1x_2x_3x_4 + x_0x_1x_2x_3 + x_0x_1 + x_0x_2 + x_0x_3 + x_0 + x_1x_4 + x_1 + x_2x_3 + x_2 + x_3x_4 + 1$ & 1777664 \\
    \hline
    288d1b41 & $x_0x_1 + x_0x_2 + x_0x_3 + x_0 + x_1x_4 + x_1 + x_2x_3 + x_2 + x_3x_4 + 1$ & 55552 \\
    \hline
    dcfdda51 & $x_0x_1x_2 + x_0x_1x_3 + x_0x_1 + x_0x_2x_4 + x_0x_3x_4 + x_0 + x_1x_2x_3 + x_1x_2x_4 + x_1x_2 + x_1x_3 + x_1x_4 + x_1 + x_2x_3 + x_3 + 1$ & 10665984 \\
    \hline
    68ad9b51 & $x_0x_1x_2 + x_0x_1 + x_0x_2x_3 + x_0x_3 + x_0 + x_1x_2x_3 + x_1x_2x_4 + x_1x_2 + x_1x_4 + x_1 + x_2x_3x_4 + x_2x_4 + x_3x_4 + 1$ & 3333120 \\
    \hline
    688ddb51 & $x_0x_1x_2x_3 + x_0x_1x_2x_4 + x_0x_1x_2 + x_0x_1 + x_0x_2x_3x_4 + x_0x_2x_3 + x_0x_2x_4 + x_0x_3 + x_0 + x_1x_2x_3x_4 + x_1x_2x_4 + x_1x_2 + x_1x_4 + x_1 + x_2x_3x_4 + x_2x_4 + x_3x_4 + 1$ & 13332480 \\
    \hline
   \caption{List of representative functions of the 48 affine equivalence classes of Boolean functions with 5 variables, with the algebraic form and the total number of functions in each class. These 48 functions were generated by~\cite{fuller2003analysis}. The algebraic normal form was obtained using sagemath. The function count was obtained by an exhaustive search.}
   \label{table_eq_classes}
\end{longtable}

\section{Cryptographic properties of 5-variable Boolean functions} \label{5_variable_properties}

\begin{longtable}{ |p{0.14\textwidth}|p{0.07\textwidth}|p{0.07\textwidth}|p{0.13\textwidth}|p{0.13\textwidth}|p{0.13\textwidth}|p{0.13\textwidth}|p{0.15\textwidth}|p{0.13\textwidth}| }
    \captionsetup{width=1.1\linewidth}
    \hline
    \textbf{Class} & \textbf{Deg.} & \textbf{Lin.} & \textbf{SAC} & \textbf{\%} & \textbf{1st ord CI} & \textbf{\%} & \textbf{Balanced} & \textbf{\%} \\
    \hline
    aa55aa55 & 1 & YES & 0 & 0 & 54 & 84.375 & 62 & 96.875 \\
    \hline
    aa55ab55 & 5 & NO & 0 & 0 & 0 & 0 & 0 & 0 \\
    \hline
    aa55bb55 & 4 & NO & 0 & 0 & 512 & 1.6129 & 15872 & 50 \\
    \hline
    aa5dbb55 & 5 & NO & 0 & 0 & 0 & 0 & 0 & 0 \\
    \hline
    aaddbb55 & 3 & NO & 0 & 0 & 16640 & 20.9677 & 59520 & 75 \\
    \hline
    aa5dbb51 & 4 & NO & 0 & 0 & 6400 & 0.288 & 833280 & 37.5 \\
    \hline
    2a5dbb51 & 5 & NO & 0 & 0 & 0 & 0 & 0 & 0 \\
    \hline
    aaddbb51 & 5 & NO & 0 & 0 & 0 & 0 & 0 & 0 \\
    \hline
    2a5dbf51 & 3 & NO & 0 & 0 & 0 & 0 & 0 & 0 \\
    \hline
    6a5dbb51 & 4 & NO & 0 & 0 & 27648 & 0.0972 & 8888320 & 31.25 \\
    \hline
    2addbb51 & 4 & NO & 0 & 0 & 76800 & 0.288 & 9999360 & 37.5 \\
    \hline
    a8ddbb51 & 4 & NO & 0 & 0 & 17920 & 1.6129 & 555520 & 50 \\
    \hline
    aeddda51 & 5 & NO & 0 & 0 & 0 & 0 & 0 & 0 \\
    \hline
    0a5dbf51 & 5 & NO & 0 & 0 & 0 & 0 & 0 & 0 \\
    \hline
    8addda51 & 5 & NO & 0 & 0 & 0 & 0 & 0 & 0 \\
    \hline
    a8dd9b51 & 5 & NO & 0 & 0 & 0 & 0 & 0 & 0 \\
    \hline
    88ddbb51 & 5 & NO & 0 & 0 & 0 & 0 & 0 & 0 \\
    \hline
    88ddbb11 & 2 & NO & 2560 & 25.8064 & 4840 & 48.7903 & 8680 & 87.5 \\
    \hline
    8c5dda51 & 3 & NO & 0 & 0 & 102912 & 0.5789 & 10554880 & 59.375 \\
    \hline
    a89d9b51 & 3 & NO & 94720 & 5.6835 & 216000 & 12.9608 & 1145760 & 68.75 \\
    \hline
    8eddda51 & 4 & NO & 0 & 0 & 7680 & 0.0072 & 23331840 & 21.875 \\
    \hline
    aefdda51 & 4 & NO & 276480 & 0.0972 & 138240 & 0.0486 & 79994880 & 28.125 \\
    \hline
    025dbf51 & 4 & NO & 645120 & 3.2258 & 17280 & 0.0864 & 6249600 & 31.25 \\
    \hline
    88ddda51 & 4 & NO & 737280 & 0.3456 & 358400 & 0.168 & 73328640 & 34.375 \\
    \hline
    88dd9b51 & 4 & NO & 163840 & 4.301 & 28160 & 0.7392 & 1666560 & 43.75 \\
    \hline
    ceddda51 & 5 & NO & 0 & 0 & 0 & 0 & 0 & 0 \\
    \hline
    0eddda51 & 5 & NO & 0 & 0 & 0 & 0 & 0 & 0 \\
    \hline
    425dbf51 & 5 & NO & 0 & 0 & 0 & 0 & 0 & 0 \\
    \hline
    8cddda51 & 5 & NO & 0 & 0 & 0 & 0 & 0 & 0 \\
    \hline
    88dddb51 & 5 & NO & 0 & 0 & 0 & 0 & 0 & 0 \\
    \hline
    289d9b51 & 5 & NO & 0 & 0 & 0 & 0 & 0 & 0 \\
    \hline
    86fdda51 & 3 & NO & 0 & 0 & 0 & 0 & 0 & 0 \\
    \hline
    88dddb71 & 3 & NO & 1310720 & 25.8064 & 0 & 0 & 0 & 0 \\
    \hline
    cefdda51 & 4 & NO & 2654208 & 1.5552 & 6144 & 0.0036 & 31997952 & 18.75 \\
    \hline
    0efdda51 & 4 & NO & 11550720 & 1.8049 & 122880 & 0.0192 & 159989760 & 25 \\
    \hline
    288d9b51 & 4 & NO & 1146880 & 4.301 & 0 & 0 & 6666240 & 25 \\
    \hline
    8cfdda51 & 4 & NO & 1966080 & 0.4608 & 414720 & 0.0972 & 133324800 & 31.25 \\
    \hline
    8cdddb51 & 4 & NO & 1904640 & 5.3571 & 102400 & 0.288 & 13332480 & 37.5 \\
    \hline
    8ccdda51 & 4 & NO & 0 & 0 & 230400 & 0.288 & 29998080 & 37.5 \\
    \hline
    289d9b41 & 4 & NO & 0 & 0 & 14336 & 1.6129 & 444416 & 50 \\
    \hline
    488ddb51 & 5 & NO & 0 & 0 & 0 & 0 & 0 & 0 \\
    \hline
    ccfdda51 & 5 & NO & 0 & 0 & 0 & 0 & 0 & 0 \\
    \hline
    688d9b51 & 5 & NO & 0 & 0 & 0 & 0 & 0 & 0 \\
    \hline
    288d9b41 & 5 & NO & 0 & 0 & 0 & 0 & 0 & 0 \\
    \hline
    288d1b41 & 2 & NO & 46592 & 83.8709 & 896 & 1.6129 & 27776 & 50 \\
    \hline
    dcfdda51 & 3 & NO & 0 & 0 & 233472 & 2.1889 & 5332992 & 50 \\
    \hline
    68ad9b51 & 3 & NO & 1582080 & 47.4654 & 69120 & 2.0737 & 1666560 & 50 \\
    \hline
    688ddb51 & 4 & NO & 3440640 & 25.8064 & 0 & 0 & 1666560 & 12.5 \\
    \hline
    \textbf{Total} & $\emptyset$ & $\emptyset$ & 27522560 & 0.6408 & 2213854 & 0.0515 & 601080390 & 13.9949 \\
    \hline
\caption{The algebraic degree and linearity (i.e. degree $\leq$ 1) for each equivalence class, as well as the number and proportion of functions in the class fulfilling the Strict Avalanche Criterion (SAC), the first order Correlation Immunity (CI) and the balancedness. We note that the total number of balanced functions is $\binom{32}{16} = 601080390$.}
\end{longtable}

\begin{longtable}{ |p{0.14\textwidth}|p{0.1\textwidth}|p{0.1\textwidth}|p{0.1\textwidth}|p{0.13\textwidth}|p{0.1\textwidth}|p{0.10\textwidth}|p{0.1\textwidth}|p{0.05\textwidth}| }
\captionsetup{width=1.1\linewidth}
    \hline
    \textbf{Class} & \textbf{PC 2} & \textbf{\%} & \textbf{PC 3} & \textbf{\%} & \textbf{PC 4} & \textbf{\%} & \textbf{PC 5} & \textbf{\%} \\
    \hline
    aa55aa55 & 0 & 0 & 0 & 0 & 0 & 0 & 0 & 0 \\
    \hline
    aa55ab55 & 0 & 0 & 0 & 0 & 0 & 0 & 0 & 0 \\
    \hline
    aa55bb55 & 0 & 0 & 0 & 0 & 0 & 0 & 0 & 0 \\
    \hline
    aa5dbb55 & 0 & 0 & 0 & 0 & 0 & 0 & 0 & 0 \\
    \hline
    aaddbb55 & 0 & 0 & 0 & 0 & 0 & 0 & 0 & 0 \\
    \hline
    aa5dbb51 & 0 & 0 & 0 & 0 & 0 & 0 & 0 & 0 \\
    \hline
    2a5dbb51 & 0 & 0 & 0 & 0 & 0 & 0 & 0 & 0 \\
    \hline
    aaddbb51 & 0 & 0 & 0 & 0 & 0 & 0 & 0 & 0 \\
    \hline
    2a5dbf51 & 0 & 0 & 0 & 0 & 0 & 0 & 0 & 0 \\
    \hline
    6a5dbb51 & 0 & 0 & 0 & 0 & 0 & 0 & 0 & 0 \\
    \hline
    2addbb51 & 0 & 0 & 0 & 0 & 0 & 0 & 0 & 0 \\
    \hline
    a8ddbb51 & 0 & 0 & 0 & 0 & 0 & 0 & 0 & 0 \\
    \hline
    aeddda51 & 0 & 0 & 0 & 0 & 0 & 0 & 0 & 0 \\
    \hline
    0a5dbf51 & 0 & 0 & 0 & 0 & 0 & 0 & 0 & 0 \\
    \hline
    8addda51 & 0 & 0 & 0 & 0 & 0 & 0 & 0 & 0 \\
    \hline
    a8dd9b51 & 0 & 0 & 0 & 0 & 0 & 0 & 0 & 0 \\
    \hline
    88ddbb51 & 0 & 0 & 0 & 0 & 0 & 0 & 0 & 0 \\
    \hline
    88ddbb11 & 0 & 0 & 0 & 0 & 0 & 0 & 0 & 0 \\
    \hline
    8c5dda51 & 0 & 0 & 0 & 0 & 0 & 0 & 0 & 0 \\
    \hline
    a89d9b51 & 0 & 0 & 0 & 0 & 0 & 0 & 0 & 0 \\
    \hline
    8eddda51 & 0 & 0 & 0 & 0 & 0 & 0 & 0 & 0 \\
    \hline
    aefdda51 & 0 & 0 & 0 & 0 & 0 & 0 & 0 & 0 \\
    \hline
    025dbf51 & 0 & 0 & 0 & 0 & 0 & 0 & 0 & 0 \\
    \hline
    88ddda51 & 0 & 0 & 0 & 0 & 0 & 0 & 0 & 0 \\
    \hline
    88dd9b51 & 0 & 0 & 0 & 0 & 0 & 0 & 0 & 0 \\
    \hline
    ceddda51 & 0 & 0 & 0 & 0 & 0 & 0 & 0 & 0 \\
    \hline
    0eddda51 & 0 & 0 & 0 & 0 & 0 & 0 & 0 & 0 \\
    \hline
    425dbf51 & 0 & 0 & 0 & 0 & 0 & 0 & 0 & 0 \\
    \hline
    8cddda51 & 0 & 0 & 0 & 0 & 0 & 0 & 0 & 0 \\
    \hline
    88dddb51 & 0 & 0 & 0 & 0 & 0 & 0 & 0 & 0 \\
    \hline
    289d9b51 & 0 & 0 & 0 & 0 & 0 & 0 & 0 & 0 \\
    \hline
    86fdda51 & 0 & 0 & 0 & 0 & 0 & 0 & 0 & 0 \\
    \hline
    88dddb71 & 0 & 0 & 0 & 0 & 0 & 0 & 0 & 0 \\
    \hline
    cefdda51 & 12288 & 0.0072 & 0 & 0 & 0 & 0 & 0 & 0 \\
    \hline
    0efdda51 & 0 & 0 & 0 & 0 & 0 & 0 & 0 & 0 \\
    \hline
    288d9b51 & 0 & 0 & 0 & 0 & 0 & 0 & 0 & 0 \\
    \hline
    8cfdda51 & 0 & 0 & 0 & 0 & 0 & 0 & 0 & 0 \\
    \hline
    8cdddb51 & 0 & 0 & 0 & 0 & 0 & 0 & 0 & 0 \\
    \hline
    8ccdda51 & 0 & 0 & 0 & 0 & 0 & 0 & 0 & 0 \\
    \hline
    289d9b41 & 0 & 0 & 0 & 0 & 0 & 0 & 0 & 0 \\
    \hline
    488ddb51 & 0 & 0 & 0 & 0 & 0 & 0 & 0 & 0 \\
    \hline
    ccfdda51 & 0 & 0 & 0 & 0 & 0 & 0 & 0 & 0 \\
    \hline
    688d9b51 & 0 & 0 & 0 & 0 & 0 & 0 & 0 & 0 \\
    \hline
    288d9b41 & 0 & 0 & 0 & 0 & 0 & 0 & 0 & 0 \\
    \hline
    288d1b41 & 28672 & 51.6129 & 10752 & 19.3548 & 1792 & 3.2258 & 0 & 0 \\
    \hline
    dcfdda51 & 0 & 0 & 0 & 0 & 0 & 0 & 0 & 0 \\
    \hline
    68ad9b51 & 199680 & 5.9907 & 0 & 0 & 0 & 0 & 0 & 0 \\
    \hline
    688ddb51 & 0 & 0 & 0 & 0 & 0 & 0 & 0 & 0 \\
    \hline
    \textbf{Total} & 240640 & 0.0056 & 10752 & 0.0002 & 1792 & 0 & 0 & 0 \\
    \hline
\caption{For each equivalence class, the number and proportion of functions in the class fulfilling the Propagation Criterion (PC) at order 2, 3, 4 and 5.}
\end{longtable}

\section{Cryptographic properties preserved after extension, by equivalence classes} \label{preserved_properties}

\begin{longtable}{ |p{0.15\textwidth}|p{0.15\textwidth}|p{0.07\textwidth}|p{0.15\textwidth}|p{0.15\textwidth}|p{0.15\textwidth}|p{0.15\textwidth}| }
    \captionsetup{width=1.1\linewidth}
    \hline
    \textbf{Class} & \textbf{SAC} & \textbf{\%} & \textbf{1st order CI} & \textbf{\%} & \textbf{Balanced} & \textbf{\%} \\
    \hline
    aa55aa55 & 0 & 0 & 54 & 84.375 & 62 & 96.875 \\
    \hline
    aa55ab55 & 0 & 0 & 0 & 0 & 0 & 0 \\
    \hline
    aa55bb55 & 0 & 0 & 512 & 1.6129 & 3978 & 12.5315 \\
    \hline
    aa5dbb55 & 0 & 0 & 0 & 0 & 0 & 0 \\
    \hline
    aaddbb55 & 0 & 0 & 4570 & 5.7585 & 17318 & 21.822 \\
    \hline
    aa5dbb51 & 0 & 0 & 0 & 0 & 42378 & 1.9071 \\
    \hline
    2a5dbb51 & 0 & 0 & 0 & 0 & 0 & 0 \\
    \hline
    aaddbb51 & 0 & 0 & 0 & 0 & 0 & 0 \\
    \hline
    2a5dbf51 & 0 & 0 & 0 & 0 & 0 & 0 \\
    \hline
    6a5dbb51 & 0 & 0 & 0 & 0 & 292406 & 1.028 \\
    \hline
    2addbb51 & 0 & 0 & 6 & 0 & 385296 & 1.4449 \\
    \hline
    a8ddbb51 & 0 & 0 & 17920 & 1.6129 & 81148 & 7.3037 \\
    \hline
    aeddda51 & 0 & 0 & 0 & 0 & 0 & 0 \\
    \hline
    0a5dbf51 & 0 & 0 & 0 & 0 & 0 & 0 \\
    \hline
    8addda51 & 0 & 0 & 0 & 0 & 0 & 0 \\
    \hline
    a8dd9b51 & 0 & 0 & 0 & 0 & 0 & 0 \\
    \hline
    88ddbb51 & 0 & 0 & 0 & 0 & 0 & 0 \\
    \hline
    88ddbb11 & 0 & 0 & 1788 & 18.0241 & 5868 & 59.1532 \\
    \hline
    8c5dda51 & 0 & 0 & 40 & 0.0002 & 442368 & 2.4884 \\
    \hline
    a89d9b51 & 0 & 0 & 28630 & 1.7179 & 155332 & 9.3205 \\
    \hline
    8eddda51 & 0 & 0 & 0 & 0 & 715958 & 0.6712 \\
    \hline
    aefdda51 & 0 & 0 & 0 & 0 & 2367024 & 0.8322 \\
    \hline
    025dbf51 & 0 & 0 & 0 & 0 & 226288 & 1.1315 \\
    \hline
    88ddda51 & 0 & 0 & 0 & 0 & 2355828 & 1.1043 \\
    \hline
    88dd9b51 & 0 & 0 & 0 & 0 & 73642 & 1.9332 \\
    \hline
    ceddda51 & 0 & 0 & 0 & 0 & 0 & 0 \\
    \hline
    0eddda51 & 0 & 0 & 0 & 0 & 0 & 0 \\
    \hline
    425dbf51 & 0 & 0 & 0 & 0 & 0 & 0 \\
    \hline
    8cddda51 & 0 & 0 & 0 & 0 & 0 & 0 \\
    \hline
    88dddb51 & 0 & 0 & 0 & 0 & 0 & 0 \\
    \hline
    289d9b51 & 0 & 0 & 0 & 0 & 0 & 0 \\
    \hline
    86fdda51 & 0 & 0 & 0 & 0 & 0 & 0 \\
    \hline
    88dddb71 & 0 & 0 & 0 & 0 & 0 & 0 \\
    \hline
    cefdda51 & 0 & 0 & 0 & 0 & 881826 & 0.5167 \\
    \hline
    0efdda51 & 0 & 0 & 0 & 0 & 4701998 & 0.7347 \\
    \hline
    288d9b51 & 0 & 0 & 0 & 0 & 232272 & 0.871 \\
    \hline
    8cfdda51 & 0 & 0 & 0 & 0 & 3816910 & 0.8946 \\
    \hline
    8cdddb51 & 0 & 0 & 0 & 0 & 420366 & 1.1823 \\
    \hline
    8ccdda51 & 0 & 0 & 0 & 0 & 967172 & 1.209 \\
    \hline
    289d9b41 & 0 & 0 & 14336 & 1.6129 & 58464 & 6.5776 \\
    \hline
    488ddb51 & 0 & 0 & 0 & 0 & 0 & 0 \\
    \hline
    ccfdda51 & 0 & 0 & 0 & 0 & 0 & 0 \\
    \hline
    688d9b51 & 0 & 0 & 0 & 0 & 0 & 0 \\
    \hline
    288d9b41 & 0 & 0 & 0 & 0 & 0 & 0 \\
    \hline
    288d1b41 & 0 & 0 & 896 & 1.6129 & 5816 & 10.4694 \\
    \hline
    dcfdda51 & 0 & 0 & 0 & 0 & 200018 & 1.8752 \\
    \hline
    68ad9b51 & 0 & 0 & 0 & 0 & 88842 & 2.6654 \\
    \hline
    688ddb51 & 0 & 0 & 0 & 0 & 48542 & 0.364 \\
    \hline
    \textbf{Total} & 0 & 0 & 68752 & 0.0016 & 18587120 & 0.4327 \\
    \hline
\caption{Number of functions and proportions of 5-variable CA rules that retain their properties of Strict Avalanche Criterion (SAC), first order Correlation Immunity and balancedness after extension to a 9-variable Boolean function, in each affine equivalence class. The proportion in \% is expressed rounded down, with 4 decimal places.}
\end{longtable}

\begin{longtable}{ |p{0.15\textwidth}|p{0.15\textwidth}|p{0.07\textwidth}|p{0.15\textwidth}|p{0.15\textwidth}|p{0.15\textwidth}|p{0.15\textwidth}| }
   \captionsetup{width=1.1\linewidth}
   \hline
   \textbf{Class} & \textbf{Prop crit 2, 3, 4, 5} & \textbf{\%} & \textbf{Deg $\geq$} & \textbf{\%} & \textbf{Nonlin.} & \textbf{\%} \\
   \hline
   aa55aa55 & 0 & 0 & 64 & 100 & 0 & 0 \\
    \hline
    aa55ab55 & 0 & 0 & 1980 & 96.6796 & 2008 & 98.0468 \\
    \hline
    aa55bb55 & 0 & 0 & 31346 & 98.7462 & 31386 & 98.8722 \\
    \hline
    aa5dbb55 & 0 & 0 & 316044 & 99.5602 & 316110 & 99.581 \\
    \hline
    aaddbb55 & 0 & 0 & 79260 & 99.8739 & 79260 & 99.8739 \\
    \hline
    aa5dbb51 & 0 & 0 & 2219978 & 99.9054 & 2219986 & 99.9057 \\
    \hline
    2a5dbb51 & 0 & 0 & 10664524 & 99.9863 & 10664614 & 99.9871 \\
    \hline
    aaddbb51 & 0 & 0 & 2221564 & 99.9767 & 2221712 & 99.9834 \\
    \hline
    2a5dbf51 & 0 & 0 & 1777656 & 99.9995 & 1777656 & 99.9995 \\
    \hline
    6a5dbb51 & 0 & 0 & 28442266 & 99.9987 & 28442266 & 99.9987 \\
    \hline
    2addbb51 & 0 & 0 & 26664524 & 99.9983 & 26664536 & 99.9984 \\
    \hline
    a8ddbb51 & 0 & 0 & 1110958 & 99.9926 & 1111006 & 99.9969 \\
    \hline
    aeddda51 & 0 & 0 & 28442612 & 99.9999 & 28442612 & 99.9999 \\
    \hline
    0a5dbf51 & 0 & 0 & 17776578 & 99.9996 & 17776626 & 99.9999 \\
    \hline
    8addda51 & 0 & 0 & 142212930 & 99.9998 & 142212966 & 99.9998 \\
    \hline
    a8dd9b51 & 0 & 0 & 26664564 & 99.9985 & 26664904 & 99.9997 \\
    \hline
    88ddbb51 & 0 & 0 & 317350 & 99.9716 & 317438 & 99.9993 \\
    \hline
    88ddbb11 & 0 & 0 & 9920 & 100 & 9920 & 100 \\
    \hline
    8c5dda51 & 0 & 0 & 17776632 & 99.9999 & 17776632 & 99.9999 \\
    \hline
    a89d9b51 & 0 & 0 & 1666556 & 99.9997 & 1666556 & 99.9997 \\
    \hline
    8eddda51 & 0 & 0 & 106659840 & 100 & 106659840 & 100 \\
    \hline
    aefdda51 & 0 & 0 & 284426232 & 99.9999 & 284426232 & 99.9999 \\
    \hline
    025dbf51 & 0 & 0 & 19998708 & 99.9999 & 19998712 & 99.9999 \\
    \hline
    88ddda51 & 0 & 0 & 213319640 & 99.9999 & 213319644 & 99.9999 \\
    \hline
    88dd9b51 & 0 & 0 & 3809264 & 99.9995 & 3809276 & 99.9998 \\
    \hline
    ceddda51 & 0 & 0 & 426639360 & 100 & 426639360 & 100 \\
    \hline
    0eddda51 & 0 & 0 & 142213116 & 99.9999 & 142213120 & 100 \\
    \hline
    425dbf51 & 0 & 0 & 319979506 & 99.9999 & 319979514 & 99.9999 \\
    \hline
    8cddda51 & 0 & 0 & 426639348 & 99.9999 & 426639360 & 100 \\
    \hline
    88dddb51 & 0 & 0 & 20316142 & 99.9999 & 20316158 & 99.9999 \\
    \hline
    289d9b51 & 0 & 0 & 26664892 & 99.9997 & 26664948 & 99.9999 \\
    \hline
    86fdda51 & 0 & 0 & 26664960 & 100 & 26664960 & 100 \\
    \hline
    88dddb71 & 0 & 0 & 5079040 & 100 & 5079040 & 100 \\
    \hline
    cefdda51 & 0 & 0 & 170655744 & 100 & 170655744 & 100 \\
    \hline
    0efdda51 & 0 & 0 & 639959040 & 100 & 639959040 & 100 \\
    \hline
    288d9b51 & 0 & 0 & 26664956 & 99.9999 & 26664960 & 100 \\
    \hline
    8cfdda51 & 0 & 0 & 426639360 & 100 & 426639360 & 100 \\
    \hline
    8cdddb51 & 0 & 0 & 35553280 & 100 & 35553280 & 100 \\
    \hline
    8ccdda51 & 0 & 0 & 79994876 & 99.9999 & 79994876 & 99.9999 \\
    \hline
    289d9b41 & 0 & 0 & 888832 & 100 & 888832 & 100 \\
    \hline
    488ddb51 & 0 & 0 & 106659840 & 100 & 106659840 & 100 \\
    \hline
    ccfdda51 & 0 & 0 & 341311488 & 100 & 341311488 & 100 \\
    \hline
    688d9b51 & 0 & 0 & 106659832 & 99.9999 & 106659840 & 100 \\
    \hline
    288d9b41 & 0 & 0 & 1777656 & 99.9995 & 1777664 & 100 \\
    \hline
    288d1b41 & 0 & 0 & 55552 & 100 & 55552 & 100 \\
    \hline
    dcfdda51 & 0 & 0 & 10665984 & 100 & 10665984 & 100 \\
    \hline
    68ad9b51 & 0 & 0 & 3333120 & 100 & 3333120 & 100 \\
    \hline
    688ddb51 & 0 & 0 & 13332480 & 100 & 13332480 & 100 \\
    \hline
\caption{Number of functions and proportions of 5-variable CA rules that retain their properties of Porpagation critera of degree 2, 3, 4 and 5, nonlinearity and keep an algebraic degree higher or equal after extension to a 9-variable Boolean function, in each affine equivalence class. The proportion in \% is expressed rounded down, with 4 decimal places.}
\end{longtable}
\end{appendices}

\end{document}